\documentclass[aps,preprint,groupedaddress,amsmath,amssymb,showpacs,floatfix]{revtex4}
\usepackage{graphicx}
\usepackage{bm}
\usepackage{epstopdf}

\providecommand{\mean}[1]{{\bm\langle} #1 {\bm\rangle}}
\providecommand{\vec}[1]{{\bm} #1 {\bm}}
\def\py{\bm{\sigma}}
\def\w{\omega}

\begin{document}

\title{Non-stationary solutions driven by thermodynamic power in the white-noise Langevin model}
\author{Mikhail P. Solon}
\email{mpsolon@gmail.com}
\affiliation{National Institute of Physics, University of the Philippines, Diliman, Quezon City, Philippines}
\author{Kristian Hauser Villegas}
\affiliation{National Institute of Physics, University of the Philippines, Diliman, Quezon City, Philippines}
\author{J.~P.~Esguerra}
\affiliation{National Institute of Physics, University of the Philippines, Diliman, Quezon City, Philippines}

\begin{abstract}
The average thermodynamic power of a time-dependent external potential in the white-noise Langevin model is derived using a Green's function solution. The power appears as a driving term in the differential equation for the average energy and determines whether the solution is stationary or non-stationary. Different dynamics are illustrated with explicit models: a linear potential with a static magnetic field, a linear potential perturbed with an oscillating component and a magnetic field switch modeled using a $\tanh$ protocol.
\end{abstract}


\maketitle
\section{Introduction}


Time-dependent potentials in the Langevin equation have been thoroughly used by Jayannavar \textit{et al.} in investigating properties of non-equilibrium systems, a few examples of which are Refs.\cite{Saha,Jayannavar,Sahoo}.  In these papers the distribution of thermodynamic work done on the Brownian particle by the external potential is analytically or numerically calculated and is then analyzed using non-equilibrium criteria such as the Jarzynski and Crook relations. The time-dependence of the potential facilitates the reversal of the protocol, which is central in comparing forward and backward work statistics. Moreover, the time-dependence sets the study of non-equilibrium statistical mechanics in the necessary context of non-equilibrium. As we will see, time-dependence is essential in producing non-stationary solutions.

Aside from the context of non-equilibrium theorems, understanding time-dependence is also interesting from a phenomenological point of view. In particular, knowledge of heavy ion and plasma thermodynamics in the presence of a time-dependent electromagnetic field could lead to methods of experimental control. An example is the controlled diffusion of overdamped Brownian particles using a time-dependent drift term \cite{Lillo}.

In this work we investigate the role of a time-dependent potential $\sim g(t) r^2$ in the white-noise Langevin model by explicitly evaluating the mean and variance of the phase space variables. We derive the formal solution using the Green's function method and pay particular attention to a differential equation describing the time evolution of the average energy. The dynamics is driven by the average thermodynamic power, which is found to be proportional to the product the time-dependent potential strength and the diffusion coefficient. Then it is the behavior of these two quantities that directly implies whether the system would be stationary or non-stationary in the long time limit. We discuss this and other physical implications by evaluating sample systems. In Sect. \ref{static} we briefly revisit a static system previously treated in Ref.\cite{liboff}. We use an alternative approach where a direct connection between the thermodynamic power and the mean behavior of the particle is established. In Sect.\ref{oscillating} we consider a linear potential perturbed with an oscillating part and we find that the calculated thermodynamic quantities exhibit steady-state oscillations as well. In Sect. \ref{switch} we treat a time-dependent magnetic field and focus on a magnetic field switch using a $\tanh$ protocol. The system is evaluated by matching perturbation solutions from the extreme values of an expansion parameter. We find that a system in equilibrium is taken out of equilibrium by the non-zero power during the time of switching and that the work done by the induced electric field is not the same for forward and backward switch protocols.

\section{Langevin model}\label{LM}
We consider a charged Brownian particle in the presence of a static magnetic field $\vec{B}=B\hat{z}$ and an external time-dependent potential $\frac{1}{2}g(t)r^2$. The Langevin equation in the $(x,y)$ plane can be written in terms of the matrix $\py=\bigl(\begin{smallmatrix}0&1\\-1&0\end{smallmatrix}\bigr)$ as
\begin{align}
\left(\begin{matrix}\dot{v}_x\\\dot{v}_y\end{matrix}\right) + \beta \left(\begin{matrix}v_x\\v_y\end{matrix}\right)-\w \py \left(\begin{matrix}v_x\\v_y\end{matrix}\right) +g(t) \left(\begin{matrix}r_x\\r_y\end{matrix}\right) = \left(\begin{matrix}\eta_x\\\eta_y\end{matrix}\right), \label{main}
\end{align}
where $\w=qB/m$ is the cyclotron frequency for a particle of charge $q$ and mass $m$, $\beta$ is the damping factor and $\vec{\eta}=\eta_x\hat{x} + \eta_y\hat{y}$ is the stochastic force per unit mass. The Cartesian components of the stochastic force are Gaussian, have zero mean and obey the autocorrelation
\begin{align}
\mean{\eta_i(t) \eta_j(t')}=\delta_{ij} \frac{kT \beta}{m} \delta(t-t').\label{autocor}
\end{align}

If we rewrite Eq.(\ref{main}) using the eigenvectors of $\py$ as a basis we can reduce our working equation to one-dimension:
\begin{align}
\dot{v}_1+ \{ \beta - i\w\} v_1+ g(t) r_1=\eta_1.\label{genmodel}
\end{align}
Recovering the real valued quantities can then be done using the definition $q_1=q_x-iq_y$ for $q \in \{v,r,\eta \}$. 

\section{Green's function solution} \label{GFS}
We solve Eq.(\ref{genmodel}) by treating the stochastic force as an inhomogeneous and finding the Green's function \cite{Dettman}. The Green's function has the properties
\begin{align}
G(t,\tau)\Big|_{\tau=t}=0 \ \ \text{and} \ \ \frac{d}{d\tau}G(t,\tau)\Big|_{\tau=t}=-1 \label{prop}
\end{align}
and obey the equation of motion
\begin{align}
\frac{d^2}{d\tau^2} G(t,\tau) + \{ \beta - i\w \} \frac{d}{d\tau} G(t,\tau) + g(\tau)G(t,\tau)=0 \label{de}.
\end{align}
From Eqs. (\ref{prop}) and (\ref{de}) the Green's function can be determined as
\begin{align}
G(t,\tau)=\frac{f_2(t)f_1(\tau)-f_1(t)f_2(\tau)}{\dot{f}_2(t)f_1(t) - f_2(t)\dot{f}_1(t)}, \label{greens}
\end{align}
where $f_1(\tau)$ and $f_2(\tau)$ are the linearly independent solutions of Eq.(\ref{de}).

Then the solution can be written as
\begin{align}
r_1=\mean{r_1} + \int_0^tG(t,\tau) \eta(\tau) d\tau, \label{pos}
\end{align}
where the ensemble average $\mean{r_1}$ is given by
\begin{align}
\mean{r_1}=\left[ v_1(0) +r_1(0)\left(\beta - i\w \right) \right]G(t,0) - r_1(0)\frac{d}{d\tau}G(t,\tau)\Big|_{\tau=0}.\label{eap}
\end{align}

The velocity is just the derivative of Eq.(\ref{pos})
\begin{align}
\dot{v_1}=\mean{v_1} + \int_0^t H(t,\tau) \eta (\tau) d\tau, \label{vel}
\end{align}
where $H(t,\tau)=\frac{d}{dt}G(t,\tau)$.

With the solutions written in this form we can use Eq.(\ref{autocor}) to arrive at the variance in position and velocity for $i=x,y$ given respectively as
\begin{align}
\mean{\Delta r_i^2} = \frac{2 kT \beta}{m} \int_0^t d\tau |G(t,\tau)|^2 \label{varp}
\end{align}
and
\begin{align}
\mean{\Delta v_i ^2} = \frac{2 kT \beta}{m} \int_0^t d\tau  |H(t,\tau)|  ^2, \label{varv}
\end{align}
where $|f|$ denotes the modulus of $f$. And we can readily compute the diffusion coefficient from the definition $D_i(t)=\frac{d}{dt} \mean{\Delta r_i^2}$ as
\begin{align}
D_i(t)=\frac{4kT \beta }{m}  \int_0^t Re \left[H(t,\tau) G^*(t,\tau) \right] d\tau,\label{dif}
\end{align}
where $f^*$ denotes the complex conjugate of $f$.

A differential equation for the time evolution of the average energy per unit mass $\mean{E}_i=\frac{1}{2}\left[  \mean{v_i^2} + \mean{\Delta v_i^2} \right]$ can be derived by multiplying Eq.(\ref{genmodel}) by $v_1^*$ and then taking the thermal average, giving
\begin{align}
\nonumber \frac{1}{2} \frac{d}{dt} \left( \mean{v_i}^2 + \mean{\Delta v_i^2}_k   \right) =  &\left( -\beta \mean{v_i}^2  - \frac{g(t)}{2}Re\left[ \mean{r_1}\mean{v_1^*} \right] \right) \\ &+  \left( - \beta \mean{\Delta v_i^2} +\frac{kT\beta}{m}+\frac{2 k T \beta g(t)}{m}   \int_0^t Re \left[ H(t,\tau) G^*(t,\tau) \right] d\tau  \right)_k \label{energy}.
\end{align}
In the derivation we used the property $H(t,t)=1$, which can be deduced from Eq.(\ref{greens}). Note that there is no magnetic field contribution in Eq.(\ref{energy}) since it cannot do work. The subscript $k$ denotes the parts of the equation which carry a factor of the Boltzmann constant $k \sim 10^{-23}$. Hence Eq.(\ref{energy}) can be seen as two equations separated by this scale. 

The macroscopic equation is
\begin{align}
\frac{1}{2} \frac{d}{dt} \mean{v_i}^2 +\beta \mean{v_i}^2 =  - \frac{g(t)}{2}Re\left[ \mean{r_1}\mean{v_1^*} \right]\label{macenergy}.
\end{align}
The term on the right hand side is the average power of the external force. Without this term the kinetic energy would vanish exponentially as $\exp[-2\beta t]$ due to damping. 

The microscopic equation,
\begin{align}
\frac{1}{2} \frac{d}{dt}  \mean{\Delta v_i^2}    +\beta \mean{\Delta v_i^2}-\frac{kT\beta}{m} =\frac{2 k T \beta g(t)}{m}   \int_0^t Re \left[ H(t,\tau) G^*(t,\tau) \right] d\tau \label{micenergy}
\end{align}
is similar except for the constant term $-\frac{kT\beta}{m}$ on the left hand side. The term on the right hand side is the average thermodynamic power. Without this term the variance in velocity would decay as $\exp[-2\beta t]$ to $\mean{\Delta v_i^2}=\frac{kT}{m}$ and the average energy per unit mass would become $\sum_i \mean{E}_i=kT/m$. Note that the power can be written as $P(t)=\frac{1}{2}g(t)D_i(t)$. This means that if the system were to equilibrate then either $g(t)$ or $D_i(t)$ should vanish at large times. If both the potential and the diffusion coefficient were non-vanishing then the variance in velocity would not be stationary in the long time limit. The behavior depends on the type of driving that $P(t)$ provides.

When the work done is performed adiabatically such that the power is negligible, we are led to a relation derived by Mazo under the assumption of an equilibrium process  \cite{Mazo}:
\begin{align}
\mean{\Delta v_i^2} = \frac{kT}{m}\left( 1-\frac{\mean{v_i(t)}^2}{v_i(0)^2} \right).
\end{align}
Clearly this does not hold in the non-adiabatic case since the work done allows for an increase in kinetic energy $\mean{v_i(t)}^2>v_i(0)^2$, which would make the variance negative.

In the following sections we look at some examples illustrating the role of the average thermodynamic power in the statistics of a Brownian particle.

\section{Static linear potential with a static magnetic field} \label{static}
When $g(t)=g$ the Green's function can be solved as
\begin{align}
G(t,\tau)=\frac{e^{\frac{\beta-i \w}{2}(\tau-t)}e^{-\frac{\mu}{2}(\tau+t)} ( e^{\mu t} - e^{\mu \tau}   )}{\mu}, \label{gconst}
\end{align}
where $\mu=\sqrt{-4g+(\beta-i\w)^2}$. With the Green's function at hand computing the thermodynamic quantities can be done following the previous analysis. However we could approach the system from a different perspective. By considering a change of variable in Eq.(\ref{de}) $\tau \to t'-s$, the equation of motion for the Green's function along a time slice $t=t'$ becomes
\begin{align}
\frac{d^2}{ds^2} G(t',t'-s) + \{ \beta-i\w \}\frac{d}{ds} G(t',t'-s) + gG(t',t'-s)=0 \label{de2}.
\end{align}
Since $s$ runs from $0$ to $t'$ it follows from the properties of $G(t,\tau)$ that the initial conditions are $G(t',t'-s)\Big|_{s=0}=0$ and $\frac{d}{ds}G(t',t'-s)\Big|_{s=0}=1$. 

Note from Eq.(\ref{genmodel}) that Eq.(\ref{de2}) is now identical in form to the equation of motion for the average position. Further more, since we are interested in the long-time behavior we can choose the initial conditions $(r_1(0),v_1(0))=(0,1)$ which correspond to $G(0,0)=0$ and $H(0,0)=1$. This reduces the average position given in Eq.(\ref{eap}) to $\mean{r_1}=G(s,0)$, which obeys the equation of motion
\begin{align}
\frac{d^2}{ds^2} G(s,0) + \{ \beta-i\w \}\frac{d}{ds} G(s,0) + g(s)G(s,0)=0.
\end{align}
Since the equations of motion and the initial conditions of $G(t',t-s)$ and $G(s,0)$ are the same, this establishes that $G(s,0)=G(t',t'-s)$ for $s \in [0,t']$. Hence we find that
\begin{align}
\nonumber \int_0^{t'} Re\left[ G^*(t',\tau)H(t',\tau) \right] d\tau&=\int_0^{t'} Re\left[G^*(t',t'-s)H(t',t'-s)\right] ds \\
\nonumber &=\int_0^{t'} Re\left[ G^*(s,0)H(s,0) \right] ds \\
\nonumber &=\int_0^{t'} \frac{1}{2} \frac{d}{ds} |G(s,0)|^2 ds \\
&=\frac{1}{2} |G(t',0)|^2. \label{connect}
\end{align}
The left hand side of the above equation is proportional to the diffusion coefficient and hence power. While the right hand side is the square of the mean position of the particle. Thus when the mean position goes to zero in the long time limit, the diffusion coefficient and power also go to zero. Consequently the variance in velocity goes to $kT/m$ according to Eq.(\ref{micenergy}). The exact behavior can be readily studied from the explicit Green's function given in Eq.(\ref{gconst}). Note that the uniqueness of the phase space equilibrium point $(\mean{r_1},\mean{v_1})=(0,0)$ guarantees that the equipartition value is the only stationary value for the velocity variance.

If we evaluate Eq.(\ref{varp}) using Eq.(\ref{gconst}) and take the long time limit, we find that the variance in position approaches
\begin{align}
\lim_{t \to \infty} \mean{\Delta r_i^2} = \frac{kT}{mg}.\label{longdeltax}
\end{align}
This defines a radius $r_0=\sqrt{\frac{kT}{mg}}$ within which the Brownian particle is most likely to be found. This result is independent of the static magnetic field, which only affects the rate at which the system diffuses to $r_0$. A stronger magnetic field results in a slower diffusion process since a spiraling path results in less displacement.

When $g<0$ the potential is repulsive and the mean becomes unbounded in phase space. Hence the diffusion coefficient and variance in velocity also increase indefinitely. 

\section{Oscillating linear potential} \label{oscillating}
We consider a system with no magnetic field $\w=0$ and an oscillating linear potential $g(t)=g + \epsilon\cos[\Omega t]$. The solutions to Eq.(\ref{de}) are
\begin{align}
f_1(\tau)=e^{\beta \tau/2}\text{MathieuC} \left[ \frac{4g-\beta^2}{\Omega^2},\frac{-2 \epsilon}{\Omega^2},\frac{g \tau}{2} \right]
\end{align}
and
\begin{align}
f_2(\tau)=e^{\beta \tau/2}\text{MathieuS}  \left[ \frac{4g-\beta^2}{\Omega^2},\frac{-2 \epsilon}{\Omega^2},\frac{g \tau}{2} \right].
\end{align}
The Green's function is then constructed using Eq.(\ref{greens}) and when $\epsilon=0$ is set, we recover the Green's function given in Eq.(\ref{gconst}) with $\w=0$.

\begin{figure}[ht]
  \begin{center}
   \includegraphics[width=0.49\linewidth,clip]{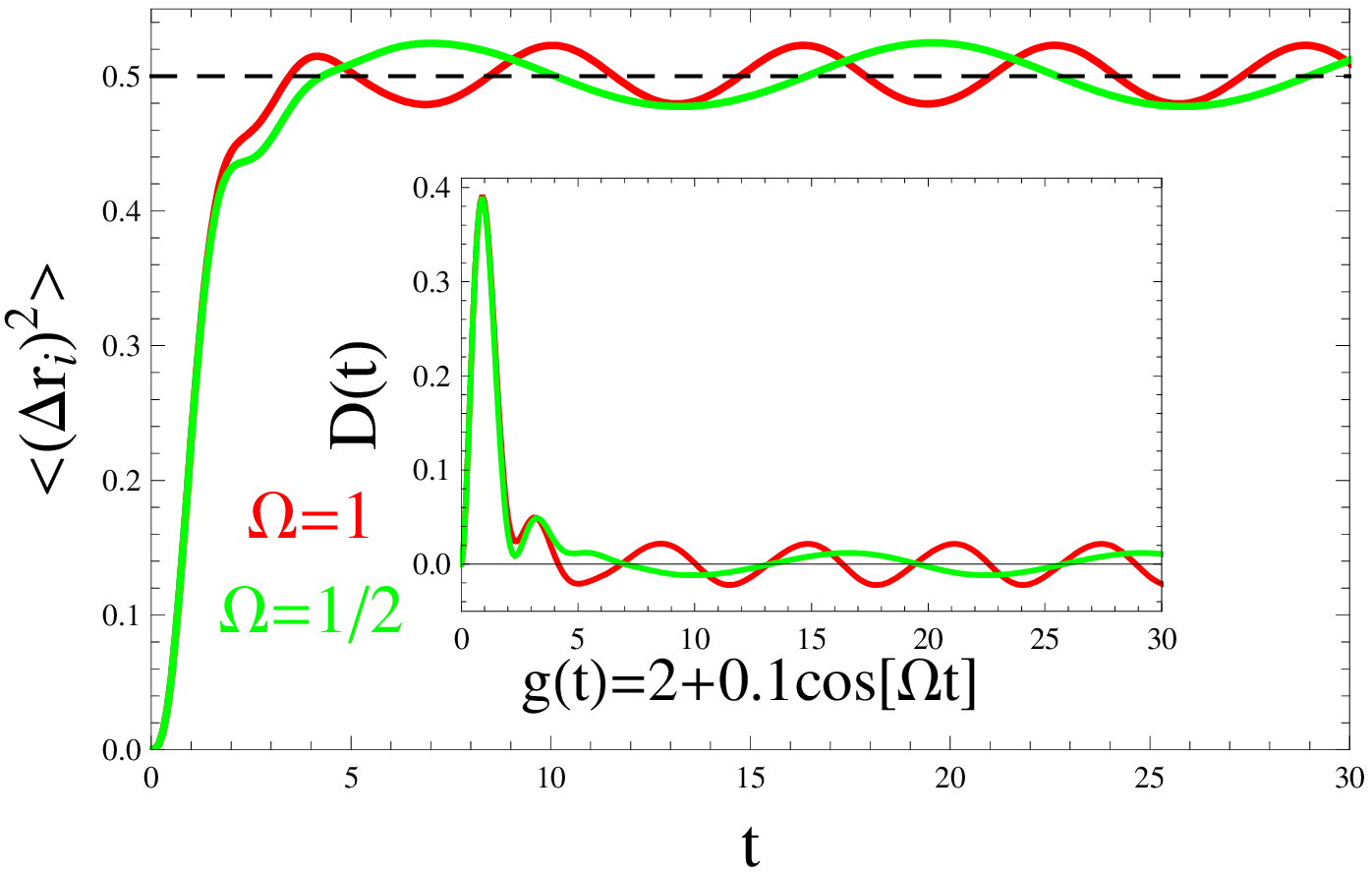}
   \includegraphics[width=0.49\linewidth,clip]{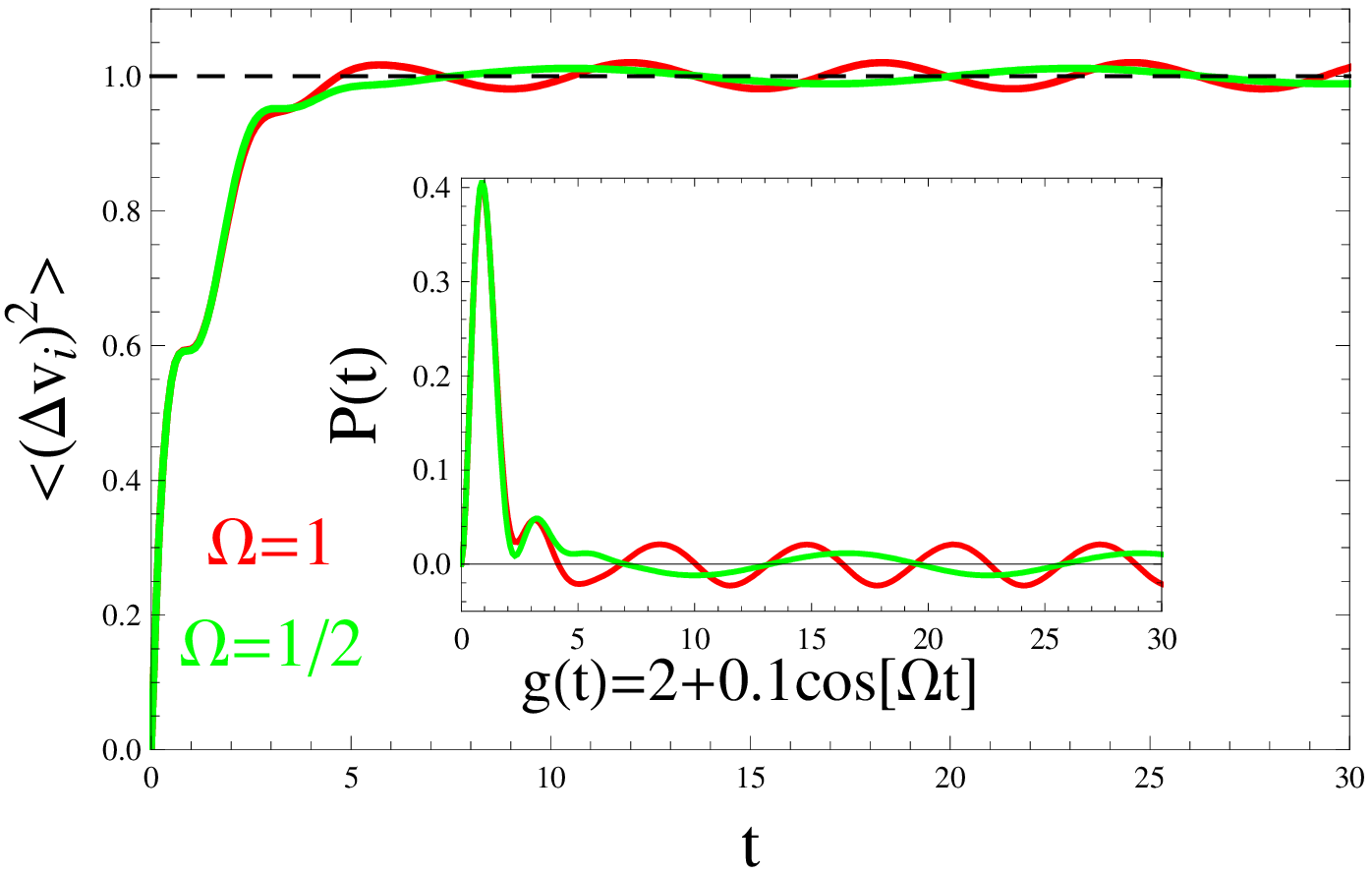}
  \end{center}
\caption{Thermodynamic quantities for an oscillating spring constant. The left (right) panel shows the variance in position (velocity) with the diffusion coefficient (power) in the inset. The black dashed line in the left (right) panel is the expected equilibrium value of $\mean{\Delta r_i^2}=1/g$ ($\mean{\Delta v_i^2}=kT/m$).\label{fig1}}
\end{figure}

If $\epsilon/g << 1$ we can consider the oscillatory part to be a perturbation and take Eq.(\ref{longdeltax}) to lowest order in $\epsilon$ as
\begin{align}
\lim_{t \to \infty} \mean{\Delta r_i^2} = \frac{kT}{mg} \left( 1-\frac{\epsilon}{g}\cos[\Omega t] \right).
\end{align}
This describes how the radius $r_0$ oscillates as it reacts to the instantaneous value of $g(t)$. Since the diffusion coefficient does not go to zero i.e. $D_i(t) \sim \Omega\sin[\Omega t]$, the thermodynamic power $P(t) \sim g(t)D_i(t)$ also remains oscillating. Then Eq.(\ref{micenergy}) implies that the variance in velocity oscillates about $kT/m$ in the long time limit. In Figure \ref{fig1} we show the numerically integrated thermodynamic quantities. In this example the harmonic strength $g(t)$ is always positive and the mean position becomes stationary at zero at large times, while the diffusion coefficient oscillates. In this case the connection between the mean dynamics and thermodynamics derived in Eq.(\ref{connect}) does not hold.

\section{Magnetic field switch} \label{switch}
In the presence of a time-dependent magnetic field $\vec{B}=B(t)\hat{z}$, an electric field of the form $\vec{E}=-\frac{r}{2}\frac{dB(t)}{dt} \hat{\phi}$ is induced. We can neglect further electromagnetic inductions because of the $\sim 1/c$ scaling. The Langevin equation we are interested in solving is 
\begin{align}
\dot{v}_1+ \{ \beta - i\w(t)\} v_1-i \lambda \frac{\dot{\w}(t)}{2}r_1=\eta_1,\label{model1B}
\end{align}
where $\lambda$ is a book-keeping parameter we will set to $1$. For some choices of $\w(t)$ the Green's function for the above equation can be solved using the previous formal solution. However, in this example we are interested in using the $\tanh$ function as a magnetic field switch protocol. We will solve for the Green's function by perturbation through constructing a matched series from the known $\lambda=0$ and $\lambda=2$ solutions.

From Eq.(\ref{model1B}) the velocity $v_1$ can be expressed implicitly as 
\begin{align}
v_1=v_1(0)\Gamma_0(t,0)+\int_0^t\Gamma_0(t,\tau)\eta_1(\tau) d\tau + i\frac{\lambda}{2}\int_0^t\Gamma_0(t,\tau)\dot{\w}(\tau)r_1(\tau)d\tau, \label{volterra}
\end{align}
where we have the velocity Green's function in the $\lambda=0$ case
\begin{align}
\Gamma_0(t,\tau)=\exp \left(-\int_{\tau}^t \left[ \beta - i\w(s)\right] ds \right).
\end{align}
To express $v_1$ explicitly as in Eq.(\ref{vel}) we iterate Eq.(\ref{volterra}), arriving at the Green's function
\begin{align}
\Gamma(t,\tau)= \sum_{p=0}^\infty \left( \frac{i \lambda}{2}\right)^p \Gamma_p(t,\tau),\label{g}
\end{align}
where
\begin{align}
\Gamma_p(t,\tau)=\int_0^tW(t,s)\Gamma_{p-1}(s,\tau)ds
\end{align}
and
\begin{align}
W(t,\tau)=\int_{\tau}^t\Gamma_0(t,s)\dot{\w}(s) ds.
\end{align}

Similarly we can construct a perturbative solution about $\lambda=2$. Note that Eq.(\ref{model1B}) can be written as
\begin{align}
\dot{v}_1+ \beta v_1 - i \frac{d}{dt} \left[\w(t) r_1(t) \right] + i \left( 2-\lambda \right) \frac{\dot{\w}(t)}{2}r_1=\eta_1. \label{model2B}
\end{align}
Thus $v_1$ can be expressed implicitly as
\begin{align}
v_1=v_1(0)\Lambda_0(t,0)+\int_0^t\Lambda_0(t,\tau)\eta_1(\tau) d\tau + i\frac{\lambda-2}{2}\int_0^t\Lambda_0(t,\tau),\dot{\w}(\tau)r_1(\tau)d\tau, \label{volterra2}
\end{align}
where we have the Green's function for the $\lambda=2$ case
\begin{align}
\Lambda_0(t,\tau)=\frac{d}{dt}\int_{\tau}^t \Gamma_0(t,s) ds.
\end{align}
Again by iteration we have the Green's function
\begin{align}
\Lambda(t,\tau)= \sum_{p=0}^\infty \left( \frac{i (\lambda-2)}{2}\right)^p \Lambda_p(t,\tau),\label{j}
\end{align}
where
\begin{align}
\Lambda_p(t,\tau)=\int_0^tX(t,s)\Lambda_{p-1}(s,\tau)ds
\end{align}
and
\begin{align}
X(t,\tau)=\int_{\tau}^t\Lambda_0(t,s)\dot{\w}(s) ds.
\end{align}

\begin{figure}[ht]
  \begin{center}
   \includegraphics[width=0.49\linewidth,clip]{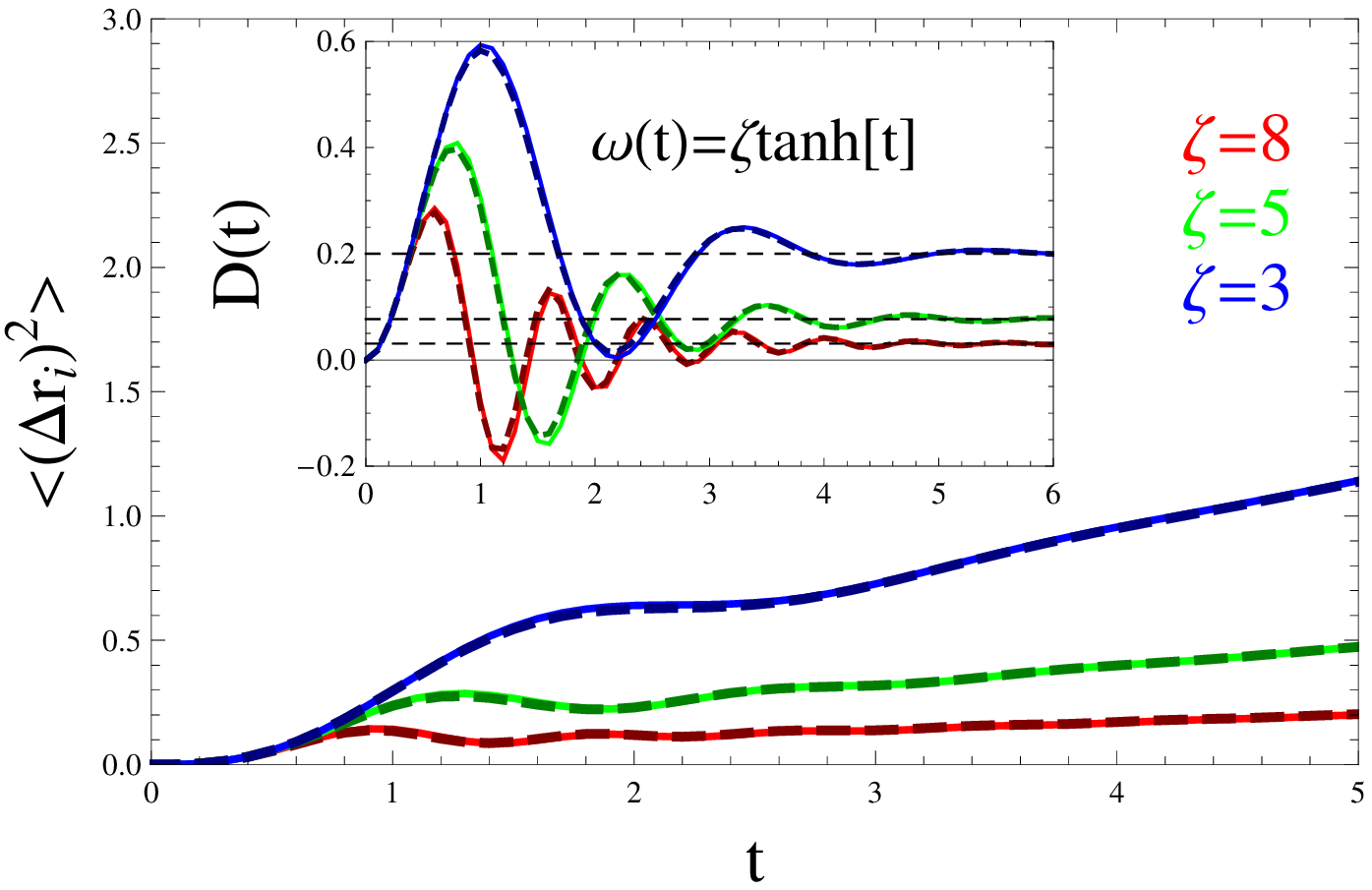}
   \includegraphics[width=0.49\linewidth,clip]{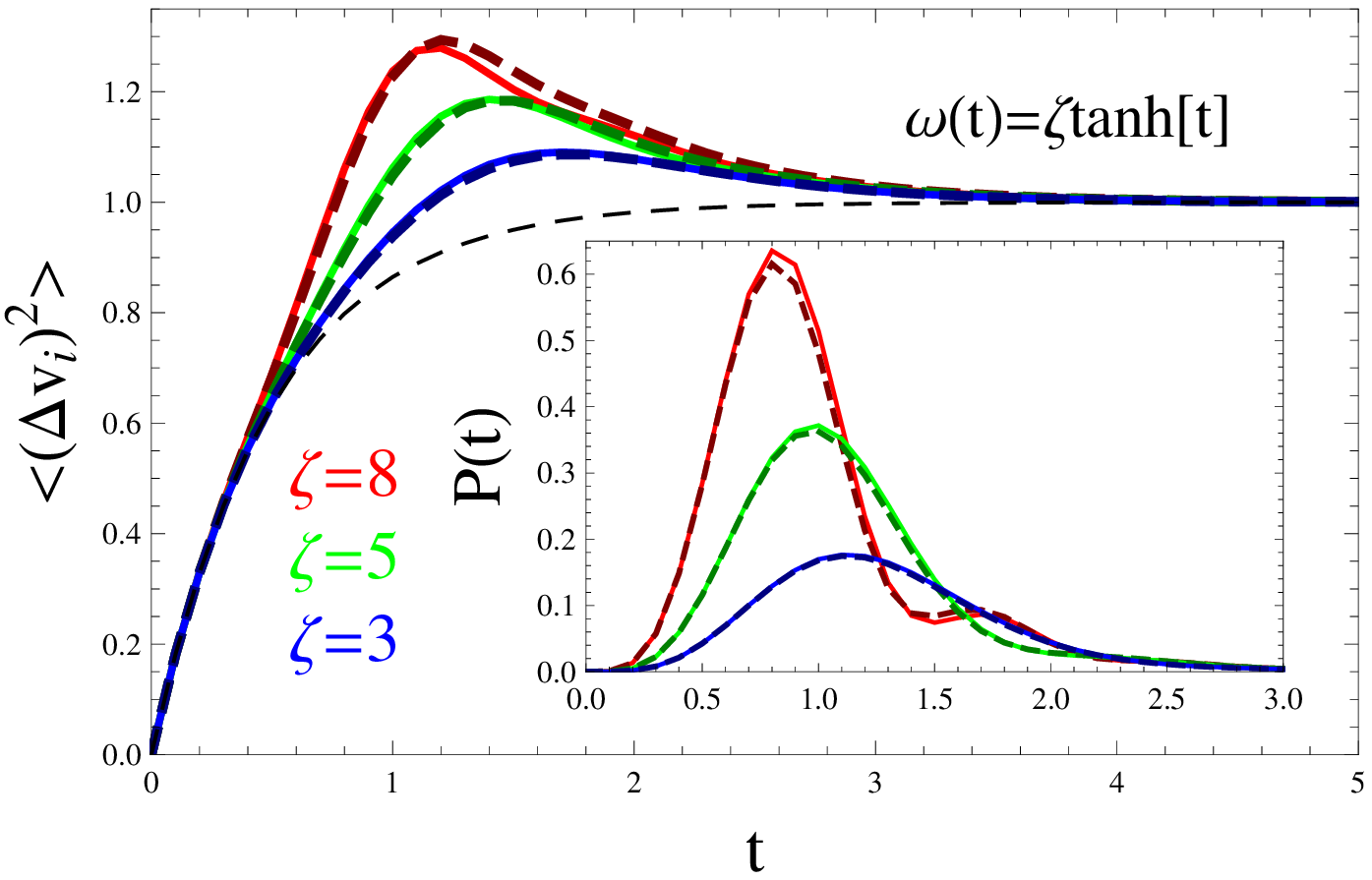}
  \end{center}
\caption{Comparison of the perturbative solution (dashed color) with the exact solution (solid color). The left (right) panel shows the variance in position (velocity) with the diffusion coefficient (power) in the inset. The black dashed lines in the diffusion coefficient inset are the values computed from Eq.(\ref{DS}). The black dashed line in the right panel is the free particle result, highlighting the $\exp[-2\beta t]$ decay.
\label{fig2}}
\end{figure}

By combining Eqs. (\ref{g}) and (\ref{j}) we connect the two solutions as
\begin{align}
H(t,\tau)=\frac{1}{2}\sum_{p=0}^\infty \left[ \left( \frac{i \lambda}{2}\right)^p \Gamma_p(t,\tau) + \left( \frac{i (\lambda-2)}{2}\right)^p \Lambda_p(t,\tau) \right].
\end{align}
We will use the zeroth order term 
\begin{align}
H_0(t,\tau)=\frac{\Gamma_0(t,\tau) + \Lambda_0(t,\tau)}{2} \label{zeroH}
\end{align}
as the velocity Green's function and
\begin{align}
G_0(t,\tau)=\int_{\tau}^t H_0(s,\tau) ds \label{zeroG}
\end{align}
as the position Green's function. To check how well the approximation performs we compare the calculations for the field $\w(t)=\xi \tanh[t]$ with those of the exact solution given by the functions
\begin{align}
f_1(\tau)=\ _2F_1\left[\frac{1-i\xi-i \sqrt{-1+\xi^2}}{2} ,\frac{1-i \xi+i \sqrt{-1+\xi^2}}{2},\frac{2-i\xi+ \beta }{2} ,\frac{1}{1+e^{2 \tau}}\right]
\end{align}
and
\begin{align}
f_2(\tau)= \ _2F_1\left[\frac{1-i \sqrt{-1+\xi^2}}{2},\frac{1+i \sqrt{-1+\xi^2}-\beta}{2},\frac{2+ i\xi -\beta}{2} ,\frac{1}{1+e^{2 t}}\right] 
\left(\frac{-1}{1+e^{2 t}}\right)^{\frac{i\xi- \beta }{2} },
\end{align}
where $ _2F_1$ is a hypergeometric function. Figure \ref{fig2} shows good agreement between the results. Note that the diffusion coefficients exhibit transient oscillations and their final values agree with the static magnetic field expression \cite{Paraan}
\begin{align}
D^{static}_i=\frac{2kT}{m}\frac{\beta}{\beta^2 + \w^2}\label{DS}.
\end{align}
The average thermodynamic power of the induced electric field is given by
\begin{align}
P(t)=Re \left[ i\dot{\w}(t) D(t) \right],
\end{align}
where $\dot{\w}(t)=\xi  \text{sech}^2 [t]$. A larger switch magnitude $\xi$ induces a stronger electric field, resulting to a larger transient spike in the velocity variance. 

Note however that the previous computations do not represent a well prepared system since the switching starts at $t=0$ when the transient exponential decay is still in effect. What we would like to see is switching in a system already in equilibrium as modeled by the protocol $\w(t)=\w_0+\xi \tanh[\sigma(t-t_s)]$ for $t_s$ larger than the exponential decay time $\frac{1}{2\beta}$. In this case the diffusion coefficient should agree with Eq.(\ref{DS}) sufficiently before and after $t_s$. Near $t_s$ the diffusion coefficient cannot be constant and the variance in position cannot be linear in time. Also the power is non-zero due to the induced electric field and the variance in velocity will be driven away from $kT/m$. 

\begin{figure}[ht]
  \begin{center}
   \includegraphics[width=0.49\linewidth,clip]{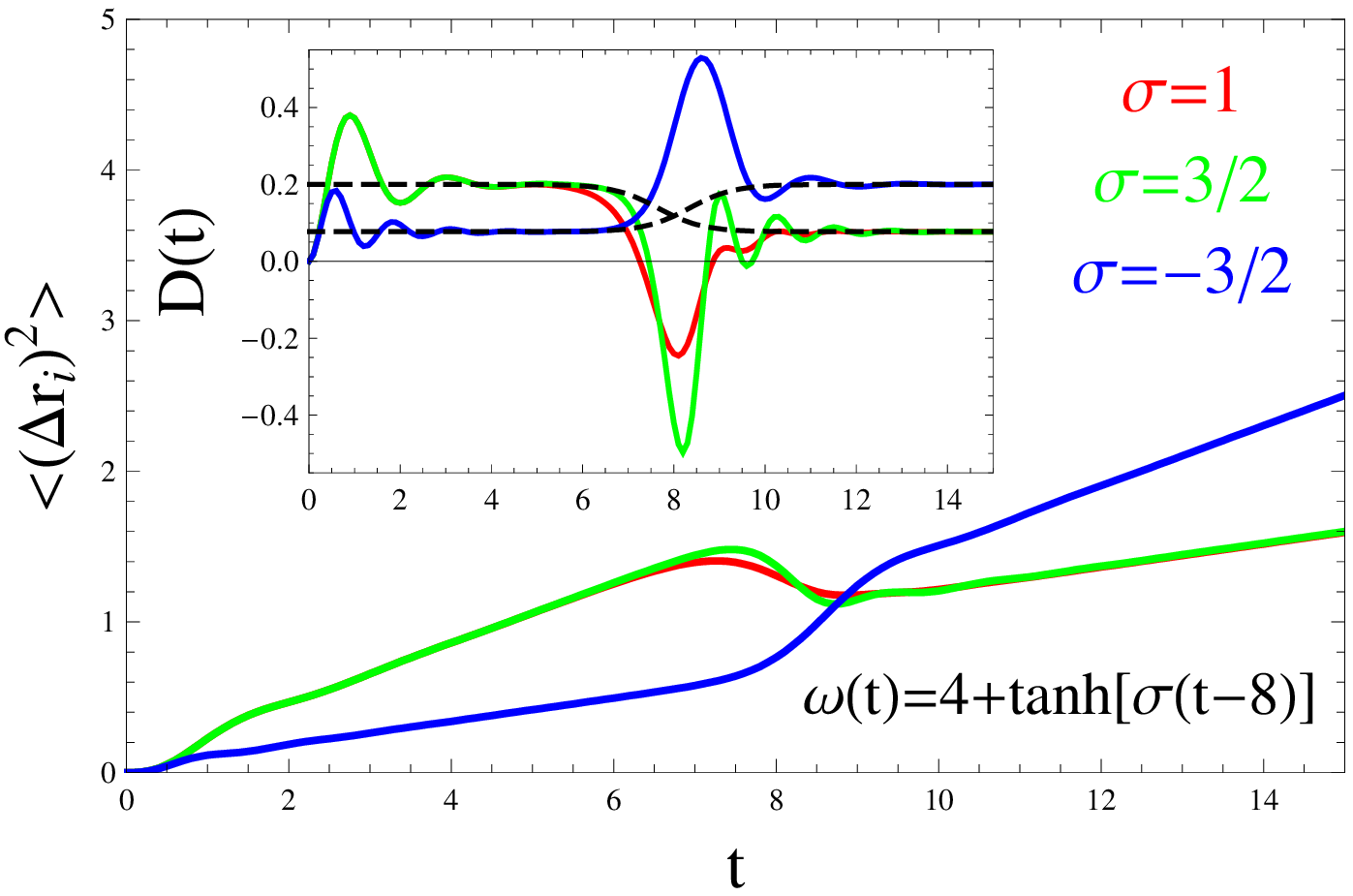}
   \includegraphics[width=0.49\linewidth,clip]{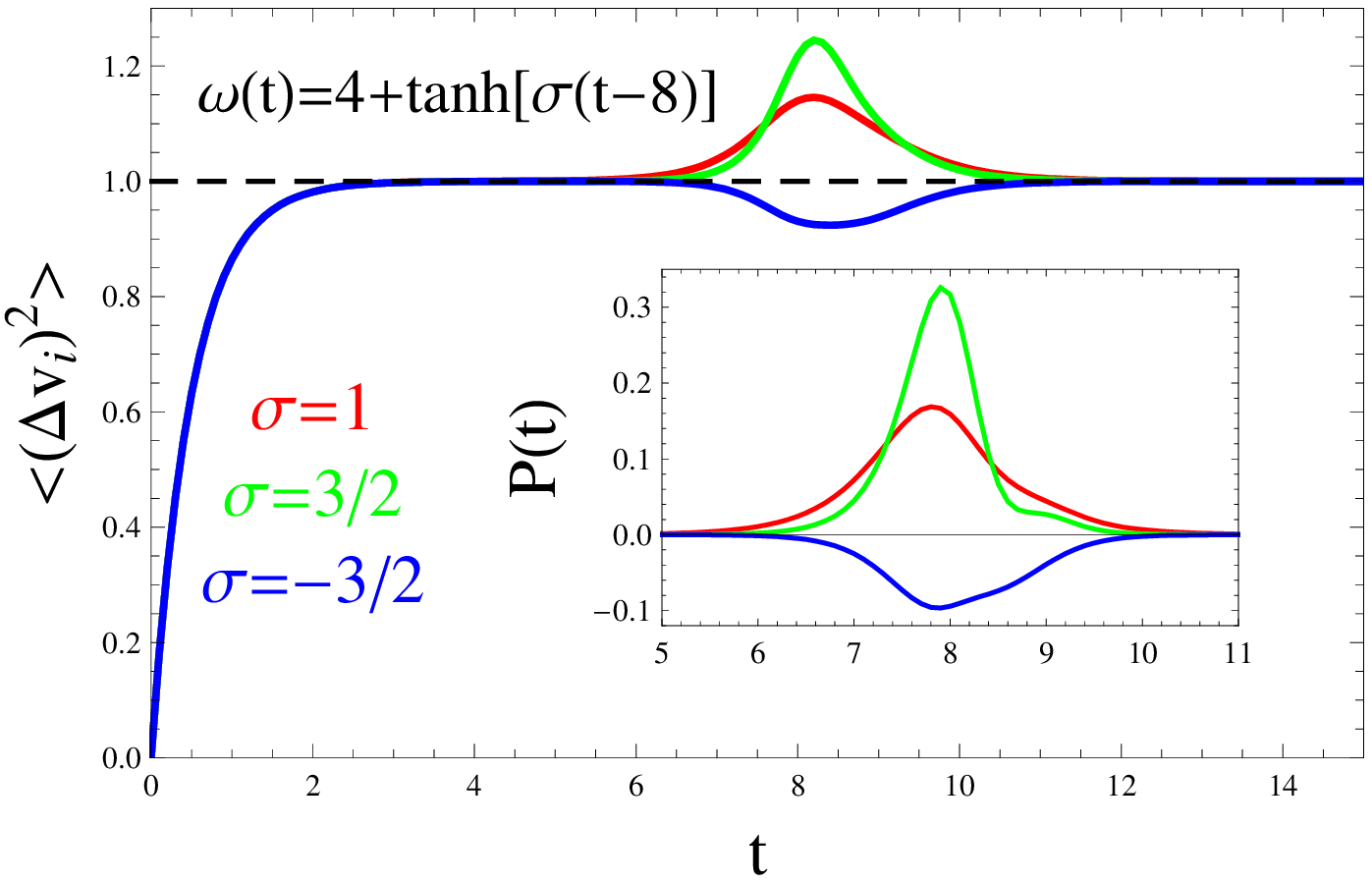}
  \end{center}
\caption{Thermodynamic quantities for a switched magnetic field. The left (right) panel shows the variance in position (velocity) with the diffusion coefficient (power) in the inset. The black dashed line in the diffusion coefficient inset is computed from Eq.(\ref{DS}), with the substitution $\w \to \w(t)$. The black dashed line in the right panel is the equipartition value $kT/m$.\label{fig3}}
\end{figure}

In Figure \ref{fig3} we show the numerically integrated thermodynamic quantities using the approximate Green's functions. Note the transient oscillations in the diffusion coefficient near $t_s$ and the expected agreement with Eq.(\ref{DS}). For the protocol considered the change in field is $\dot{\w}(t)=\sigma \text{sech}^2 [\sigma(t-t_s)]$. The power is positive when $\sigma>0$ and the work done by the induced field causes the variance in velocity to spike above the equipartition value. Moreover, increasing the rate of the switch $\sigma$ induces a stronger electric field, resulting in a larger spike. Thus a slower rate corresponds to a smaller spike, which agrees with the expectation that in the adiabatic limit the system would  be kept at equilibrium. When $\sigma<0$ the power is negative and the variance in velocity dips below the equipartition value. 

Since $\tanh$ is an odd function, the protocols for $\sigma_{forward}=3/2$ and $\sigma_{backward}=-3/2$ are time-reversed with respect to each other. While the initial and final values of the diffusion coefficients match accordingly, it is evident from the figure that the transient process is not the same for the forward and backward protocols. There is a difference in the average power which is in fact the basis of testing non-equilibrium work theorems as in Ref.\cite{Saha}.

\section{Conclusion} \label{conclusions}
A time-dependent Langevin system may exhibit a stationary or non-stationary solution depending on the driving provided by the thermodynamic power $P(t)=\frac{1}{2}g(t)D(t)$. Equilibration occurs when the power vanishes, implying that the variance in velocity takes the equipartition value. This happens if $g(t)$ goes to zero at large times as in the case of the magnetic field switch, or if the diffusion coefficient goes to zero as in the case of a static linear potential. Non-equilibration happens when the power does not vanish in the long-time limit. The power could oscillate as in the case of a linear potential with a perturbative oscillating component, resulting in to thermodynamic quantities that oscillate as well. The power could also increase indefinitely, for example when $g(t)=g<0$, resulting to unbounded variances.

\newpage

\end{document}